\documentclass[aps,prx,superscriptaddress,showpacs,twocolumn]{revtex4-2}
\usepackage{graphicx}% Include figure files
\usepackage{dcolumn}% Align table columns on decimal point
\usepackage{bm}% bold math
\usepackage{color}
\usepackage[table]{xcolor}
\usepackage{amsmath}
\usepackage{amssymb}
\usepackage[caption=false]{subfig}
\usepackage{amsmath}
\usepackage{float}
\usepackage{natmove}
\usepackage{multirow}
\usepackage{setspace}
\usepackage{array}
\usepackage{booktabs}
\usepackage{rotating}
\usepackage{comment}
\usepackage{xr-hyper}
\makeatletter
\newcommand*{\addFileDependency}[1]{% argument=file name and extension
  \typeout{(#1)}
  \@addtofilelist{#1}
  \IfFileExists{#1}{}{\typeout{No file #1.}}
}
\makeatother

\usepackage[colorlinks=true,linkcolor=blue,citecolor=blue,urlcolor=blue,filecolor=blue]{hyperref}
\usepackage{physics}
\usepackage{mathrsfs}
\usepackage{tabularx}
\usepackage{layouts}
\usepackage{ulem}
\setcitestyle{numbers,square}

\newcommand{\Trw}{\text{Tr}_\omega}

\newcolumntype{C}{>{\centering\arraybackslash}X}

\def\QE{\textsc{Quantum ESPRESSO}\,}

\definecolor{tangerine}{rgb}{0.944,0.522,0}
\definecolor{brown}{rgb}{0.633,0.156,0.156}
\definecolor{lime}{rgb}{0.5,1.0,0.0313}
\definecolor{limedark}{rgb}{0.333, 0.666, 0.020}
\definecolor{applegreen}{rgb}{0.55, 0.71, 0.0}
\definecolor{green1}{rgb}{0.0, 0.5, 0.0}
\definecolor{green2}{rgb}{0.25, 0.5, 0.016}
\definecolor{BluBondi}{rgb}{0.00,0.58,0.71}
\definecolor{myred}{rgb}{0.784, 0.063, 0.180}  % 200 16 46
\definecolor{mygreen}{rgb}{0.478,0.604,0.004}  % 122 154 1
\definecolor{myblue}{rgb}{0.059,0.298,0.506}   % 15 76 129
\newcommand{\editor}[2]{%
  \expandafter\newcommand\csname #1note\endcsname[1]{%
    \textcolor{#2}{(\textbf{#1:} {\it ##1})}}%
  \expandafter\newcommand\csname #1\endcsname[1]{%
    \textcolor{#2}{##1}}%
  \expandafter\newcommand\csname #1cancel\endcsname[1]{%
    \textcolor{#2}{\sout{##1}}}%
  \expandafter\newcommand\csname #1change\endcsname[2]{%
    \textcolor{#2}{\sout{##1} ##2}}%
  \newenvironment{#1text}{\color{#2}}{\color{black}}
}

\editor{AF}{limedark}
\editor{NM}{green}
\editor{TC}{purple}
\editor{AP}{tangerine}
\editor{MQ}{green2}
\editor{MC}{cyan}
\editor{note}{BluBondi}

\newcommand{\epfl}{Theory and Simulation of Materials (THEOS), and National Centre for Computational Design and Discovery of Novel Materials (MARVEL), \'Ecole Polytechnique F\'ed\'erale de Lausanne, CH-1015 Lausanne, Switzerland}

\newcommand{\piesai}{PSI Center for Scientific Computing, Theory and Data, 5232 Villigen PSI, Switzerland}   

\newcommand{\caltech}{Department of Applied Physics and Materials Science, California Institute of Technology, Pasadena, California 91125, USA}    

\newcommand{\cnr}{Centro S3, CNR--Istituto Nanoscienze, 41125 Modena, Italy}
\begin{document}
\title{Self-consistent dynamical Hubbard functional for correlated solids}

%%
%% AUTHORS
%% tentative order, to be rationalized
%%
\author{Tommaso \surname{Chiarotti}}
\email[corresponding author: ]{chiarott@caltech.edu}
\affiliation{\caltech}
\author{Matteo \surname{Quinzi}}
\affiliation{\epfl}
\author{Andrea \surname{Pintus}}
\affiliation{\epfl}
\author{Mario \surname{Caserta}}
\affiliation{\epfl}
\author{Andrea \surname{Ferretti}}
\affiliation{\cnr}
\author{Nicola \surname{Marzari}}
\affiliation{\epfl}
\affiliation{\piesai}

% PACS
%
\date{\today}

%
% ABSTRACT
%
% AF: abstract just tentative, feel free to modify at large
%
\begin{abstract}
Many-body functionals of the Green's function can provide fundamental advances in electronic-structure calculations, due to their ability to accurately predict both spectral and thermodynamic properties—such as angle-resolved photoemission spectroscopy (ARPES) experiments and total energies of materials.  
However, fully first-principles, self-consistent calculations with these dynamical functionals remain a major challenge, ultimately limiting their application to thermodynamic quantities, and restricting spectral predictions to one-shot calculations.  
In this paper, we present a fully self-consistent treatment of the electronic structure of solids using a dynamical functional.  
Our approach leverages the so-called dynamical Hubbard functional, which generalizes the DFT+$U$ correction by incorporating frequency-dependent screening, augmenting the static density functional to accurately describe both spectral and thermodynamic properties of materials with $d$- or $f$-localized orbitals near the Fermi level. 
To enable this, we employ the algorithmic-inversion method based on a sum-over-poles representation, resulting in a numerically accurate self-consistent scheme for frequency-integrated properties, while keeping real-axis spectral resolution for dynamically-resolved quantities.
Using this framework, we study the paradigmatic correlated solid SrVO$_3$, accurately reproducing its spectral features—essentially confirming previous one-shot predictions—and improving the description of its equilibrium properties, such as the equilibrium volume and bulk modulus, bringing these significantly closer to experimental measurements.
\end{abstract}

%
%%%%%%%%%%%%%%%%%%%%%%%%%
% Main body
%%%%%%%%%%%%%%%%%%%%%%%%%
%
\maketitle
%
% comment the line below to get rid of TOC
% \tableofcontents

% \input{sec_main.tex}
\section{Introduction}
\label{sec:intro}
First-principles calculations are among the most important tools for the understanding, design, and discovery of materials~\cite{marzari_electronic-structure_2021}.
These calculations are routinely performed using Kohn-Sham (KS) density-functional theory (DFT) to predict thermodynamic properties such as equilibrium structures, bulk moduli, lattice dynamics, and formation energies are most often augmented by dynamical methods like $GW$ or dynamical mean-field theory when addressing spectral properties, such as band structures and satellites~\cite{onida_electronic_2002,georges_dynamical_1996}.
A spectral functional theory that can accurately capture both thermodynamic and spectral properties---especially for correlated materials---remains a fundamental challenge in the current electronic-structure toolbox. 
Such a theory would significantly advance first-principles predictions, enabling the consistent treatment of dynamical correlations (such as band renormalization and spectral weight transfer) alongside thermodynamic quantities.
Dynamical energy functionals of the Green's function offer this flexibility, as they are, in principle, able to accurately predict both spectral and thermodynamic properties of materials. 
Although originally introduced by Luttinger, Ward, and Klein~\cite{luttinger_ground-state_1960,baym_conservation_1961,baym_self-consistent_1962,martin_interacting_2016,stefanucci_nonequilibrium_2013}, the approximations required for the so-called $\Phi_{xc}$ exchange-correlation functional, along with the added complexity introduced by their dynamical nature, have so far limited their widespread use and adoption.

% Diagrammatic functionals of many-body perturbation theory are a paradigmatic example of this, where to go beyond $GW$ has been proven problematic both for numerical and conceptual issues~\cite{}.
The $GW$ approach is a paradigmatic example of both advantages and limitations.  
Although it is an established and fundamental tool for predicting spectral properties of materials~\cite{reining_gw_2018,martin_interacting_2016}, calculations are most often carried out non-self-consistently ($G_0W_0$), making the results dependent on the starting DFT ground state. 
This dependence becomes particularly critical when DFT fails dramatically in its predictions, such as in the case of incorrect metallic behavior or erroneous magnetic ground states.
This can be addressed using quasiparticle self-consistent $GW$ calculations (QPSCGW)~\cite{vanschilfgaarde_quasiparticle_2006,kotani_quasiparticle_2007}, though not including and weighting correctly satellites and spectral-weight transfers.
Despite being a functional approach, fully self-consistent ab-initio $GW$ calculations for materials are not routinely performed, with some exceptions based on imaginary-axis formulations~\cite{kutepov_ground-state_2009,kutepov_electronic_2012,grumet_beyond_2018,yeh_fully_2022,iskakov_greenweakcoupling_2025}.
In these approaches, dynamical operators are sampled on the imaginary axis, which greatly improves the numerical accuracy required for self-consistent calculations.
Though powerful, challenges emerge when dynamical quantities are analytically continued to the real axis to obtain spectral properties~\cite{gunnarsson_analytical_2010,kutepov_one-electron_2017}.
Although recent advancements have been proposed to address this~\cite{fei_nevanlinna_2021,zhang_minimal_2024}, achieving accurate spectral and thermodynamic properties using imaginary-axis methods remains challenging---especially for quasiparticles and satellites located several eV away from the Fermi level, where the distance from the imaginary axis becomes significant.

Dynamical mean-field theory (DMFT) offers another possible route to address spectral quantities for correlated materials~\cite{georges_dynamical_1996,kotliar_electronic_2006}.  
In addition to being highly successful in solving the Hubbard model early on, the combination of DFT and DMFT---where one self-consistently builds, from the charge density, an effective Hubbard model and solves it---can be viewed as a functional theory of the Green's function and allows to treat real materials~\cite{savrasov_spectral_2004,kotliar_electronic_2006}.
Though very powerful, the need to solve a many-body impurity model limits the size of the correlated manifold (for a review on the scaling of different solvers see, e.g., Ref~\cite{pavarini_dynamical_2022}). 
Furthermore, the choice of the interaction parameter $U$ in the effective model can be challenging, with predictions sometimes depending on its value~\cite{PhysRevB.97.125120,PhysRevResearch.6.033122,PhysRevB.102.045146}.
The extension of DMFT to include dynamical screening in combination with the $GW$ method, GW+EDMFT, removes the dependence on the chosen $U$~\cite{boehnke_when_2016,nilsson_multitier_2017}.
However, this comes at the cost of increased complexity in an already computationally demanding scheme.
Within this framework, solvers routinely rely on imaginary-axis sampling of dynamical propagators, leading to similar issues as the $GW$ methods discussed above.

Here, we propose a different route to self-consistently address the electronic structure of correlated solids from first principles.  
Namely, we add a (semi)local and dynamical term to DFT, effectively introducing a novel spectral functional of the Green's function.  
As detailed in Sec.~\ref{sec:dynH}, this functional approach generalizes the rotationally invariant +U correction to DFT by Dudarev et al.~\cite{dudarev_electron-energy-loss_1998} to include dynamical screening.  
In Refs.~\cite{chiarotti_energies_2024,caserta_dynamical_2025}, we introduced and applied this method---termed DFT+dynamical Hubbard (DFT+dynH)---to compute the electronic structure of the paradigmatic cubic perovskite SrVO$_3$ and four charge-transfer/Mott-Hubbard monoxides MnO, FeO, CoO, and NiO, obtaining excellent agreement with experiments.

In this work, we introduce a fully self-consistent framework for dynamical Hubbard calculations and use it to study the spectral and thermodynamic properties of SrVO$_3$.  
Specifically, we develop both charge- and fully self-consistent methodologies to find the stationary point of the dynamical Hubbard functional in an ab-initio framework.  
Crucially, the algorithmic-inversion method on sum over poles (AIM-SOP) allows to faithfully represent the frequency dependence of the propagators on the whole complex plane and solve the related Dyson equations, thereby avoiding analytic continuation procedures.
Furthermore, the framework ensures the numerical accuracy required for self-consistency, similarly to DFT implementations.
As detailed in Sec.~\ref{sec:AIM-SOP}, within the AIM-SOP method, propagators are expanded onto SOP basis, leveraging the explicit knowledge of their poles and residues.  
This allows to build a non-interacting system with additional degrees of freedom (DOF), which exactly reproduces the Green's function of the interacting system when the augmented DOF are projected away.
For the case of SrVO$_3$, the resulting self-consistent spectral function remains very similar to our previous one-shot calculations~\cite{chiarotti_energies_2024}, confirming earlier results from self-consistent quasiparticle $GW$ calculations on the same material~\cite{gatti_dynamical_2013}. 
However, the improvements in both the equilibrium lattice parameter and bulk modulus highlights the importance of self-consistency for accurately capturing thermodynamic properties.

% Furthermore, the lowering of the bulk modulus is confirmed (and even enhanced) by self consistency, showing dynamical correlation can play an essential for the equilibrium properties of these perovskites.

% This paper is organized as follows:

\section{Method}
\label{sec:method}
In this Section, we provide a review of the formalism, including the dynamical Hubbard functional together with the AIM-SOP method, as the cornerstones of the self-consistent formulation used in this work.  
We then present the self-consistent methodology, first for charge self-consistency and subsequently for fully self-consistent calculations.

\subsection{Dynamical Hubbard functional}
\label{sec:dynH}
Here, we briefly review the dynamical Hubbard functional; 
one can refer to Refs.~\cite{chiarotti_energies_2024,chiarotti_spectral_2023} for a more detailed discussion.  
As mentioned in the Introduction, the dynamical Hubbard functional is a spectral (Klein) functional of the Green's function that generalizes the rotationally invariant +U correction in DFT to include dynamical screening~\cite{dudarev_electron-energy-loss_1998}.  
This energy functional reads:
\begin{eqnarray}
\label{eq:klein_locGW}
   E_{\text{dynH}}[{G}] &=& E_H[\rho] + E_{xc}[\rho] + \Phi_{\text{dynH}}[\mathbf{G}] 
   \\[6pt] \nonumber 
   &+& \text{Tr}_\omega \left[I - {G}_0^{-1} {G} \right] 
   + \text{Tr}_\omega \log (G_0^{-1} G) + \text{Tr}_\omega [h_0 G_0].
\end{eqnarray}
Here, $G$ is the Green's function (GF) of the system, ${G}_0^{-1} = \omega I - h_0$ represents the non-interacting Green's function with $h_0 = T + v_\text{ext}$ ($v_\text{ext}$ is the external pseudo potential), $\rho$ is the density derived from ${G}$, $\mathbf{G}$ is the projection of $G$ onto a localized manifold---referred to as the Hubbard manifold and typically spanning $d$- or $f$-states---,  
and $\text{Tr}_\omega [...]$ stands for $\int \frac{d\omega}{2 \pi i} e^{i\omega 0^+} \text{Tr}[...]$.

In its most general multi-site, spin-resolved form, the dynamical Hubbard exchange-correlation $\Phi_{xc}$ term is given by~\cite{caserta_dynamical_2025}:
\begin{equation}
    \begin{split}
      \Phi_\text{dynH}[\mathbf{G}] =  &
      \frac{1}{2} \int \frac{d\omega}{2\pi i}  e^{i\omega 0^+} \sum_{\sigma,I} 
      \Bigg(U_{I}^{\infty} \ \text{Tr}[\mathbf{G}_I^\sigma(\omega)] \\
      - & \int  \frac{d\omega'}{2\pi i} e^{i\omega' 0^+} U_I(\omega') 
            \text{Tr}[\mathbf{G}_I^\sigma(\omega+\omega') \mathbf{G}_I^\sigma(\omega)]\Bigg),
    \end{split}
    \label{eq:dynH_functional}
\end{equation}
with $U(\omega)$ the average of the RPA-screened Coulomb potential over the local manifold (see also the Supplemental Material of Ref.~\cite{chiarotti_energies_2024}).  
Adopting the same approximation of Refs.~\cite{chiarotti_energies_2024,caserta_dynamical_2025}, the first term on the right-hand side of Eq.~\eqref{eq:dynH_functional} represents a double-counting correction in the fully localized limit.

For completeness, we also report the self-energy per site and spin channel derived from $\Phi_\text{dynH}$:
\begin{equation}
\begin{split}
    \mathbf{\Sigma}_\text{dynH}^{\sigma,I}(\omega) =
     -\int \frac{d\omega'}{2\pi i} e^{i\omega' 0^+} U_I(\omega')\mathbf{G}_I^{\sigma}(\omega + \omega')  +  \frac{U_{I}^{\infty}}{2}.
\end{split}    
\label{eq:dynHUB_SE}
\end{equation}
As mentioned above, this is a multi-site, spin-resolved version of the dynamical Hubbard functional introduced in Ref.~\cite{chiarotti_energies_2024}.  
The choice of the underlying DFT approximation is important, as it affects the value of the RPA-screened averaged Coulomb potential $U(\omega)$, particularly in solids that are incorrectly predicted to be metallic by local or semi-local DFT (e.g., FeO or CoO);
see Ref.~\cite{caserta_dynamical_2025} for further discussions.
The localized Hubbard manifold can be constructed from, e.g., maximally localized Wannier functions~\cite{marzari_maximally_2012}, or from (ortho-)atomic projectors;  
in this work, we use ortho-atomic projectors, as in Ref.~\cite{chiarotti_energies_2024}.

Finally, the $k$-dependent self-energy entering the Dyson equation reads:
\begin{equation}
    {\Sigma}_{\text{dynH}}^{\sigma}(\omega)=\sum_{m m',I,\mathbf{R}} \ket{\phi_{m,\mathbf{R}}^I} \mathbf{\Sigma}_{\text{dynH}}^{m m',I,\sigma}(\omega) \bra{\phi_{m',\mathbf{R}}^I},
    \label{eq:unfolding_SE}
\end{equation}
where $\ket{\phi_{m,\mathbf{R}}^I}$ are the above mentioned local (Wannier-like) orbitals on site $I$ and in cell $\mathbf{R}$.
\begin{comment}
    \TCnote{check this last equation}\MQnote{Should we add the k-dependent expression by adding the identity over KS states? i.e., removing summation over $\mathbf{R}$ and adding the identity in KS space, here below}
\begin{align}
    {\Sigma}_{\text{dynH}}^{\sigma}(\omega,\mathbf{k}) &= \sum_{n n'}\sum_{m m',I}\ket{\psi_n(\mathbf{k})} \langle{\psi_n(\mathbf{k})|\phi_{m}^{I}}\rangle\\ &\mathbf{\Sigma}_{\text{dynH}}^{m m',I}(\omega) \langle{\phi_{m'}^{I}|\psi_{n'}(\mathbf{k})}\rangle\bra{\psi_{n'}(\mathbf{k})}    
\end{align}
\end{comment}

\subsection{Algorithmic-inversion method on sum over poles}
\label{sec:AIM-SOP}
The algorithmic-inversion method on sum over poles (AIM-SOP) is a theoretical and computational framework tailored for dynamical formulations from first principles.  
The method was introduced in Ref.~\cite{chiarotti_unified_2022} for scalar propagators and applied to the homogeneous electron gas, and later generalized to the operatorial case (required for realistic materials) in Ref.~\cite{chiarotti_energies_2024,chiarotti_spectral_2023}; further details and applications can also be found in Refs.~\cite{ferretti_greens_2024,quinzi_broken_2025}.
It consists in expanding all dynamical propagators on a sum-over-poles (SOP) basis.  
For example, this expansion for the self-energy (in real-space degrees of freedom) reads:
\begin{equation}
    \Sigma(\mathbf{r},\mathbf{r'},\omega) = \sum_{i=1}^{N} \frac{\Gamma_i(\mathbf{r},\mathbf{r'})}{\omega-\Omega_i} + \Sigma_0(\mathbf{r},\mathbf{r'}),
    \label{eq:SOP_SE}
\end{equation}
where $\Gamma_i$ are operatorial residues, $\Omega_i$ are scalar poles, and $\Sigma_0$ is a static operator (e.g., the Hartree-Fock term).  
In general, all time-ordered (TO) operators can be expanded in SOP form~\cite{engel_calculation_1991}.  
Moreover, operations such as addition, multiplication, and convolution are closed on SOP; i.e., given two propagators expanded on SOP, the result of these operations is again a SOP, with residues and poles expressed as analytic functions of the inputs~\cite{chiarotti_unified_2022}.
%\footnote{Note that when convolving time-ordered propagators, the chemical potential of the resulting SOP is given by the difference of the chemical potentials of the input SOPs.}
The crucial aspect of the SOP representation is in the solution of the Dyson equation (below we omit spatial indexes for simplicity):
\begin{equation}
    G(\omega) = [\omega - h_0 - \Sigma(\omega)]^{-1}
    \label{eq:dyson_eq}
\end{equation}
via a mapping to a fictitious noninteracting system with augmented degrees of freedom, constrained to reproduce the same Green's function as the interacting system upon projection.
Given the factorization of the residues $\Gamma_i=V_m\bar{V}^\dagger_m$~\footnote{
Here, any singular value decomposition suffice to find $V_m$ and $\bar{V}^\dagger_m$.
In this work we use $\Gamma_i=\sqrt{\Gamma_i}\sqrt{\Gamma_i}$ as defined by (omitting the bath index for brevity): $\Gamma = S \sqrt{\gamma} \sqrt{\gamma} S^{-1} $, with $\sqrt{\gamma}$ the diagonal matrix of the (principal branch) of the square root of the eigenvalues.
}, to exactly mimic the effect of the self-energy, we couple the noninteracting Hamiltonian $h_0$ to $N$ fictitious noninteracting baths via $V_m$ ($\bar{V}^\dagger_m$) and having energies $\Omega_i$. It can be shown (see below) that once the fictitious degrees of freedom (DOFs) are projected out, the noninteracting auxiliary system (AIM) and the real interacting system share the same Green's function.  

The effective Hamiltonian of the AIM system is:
\begin{equation}
    {H}_\text{AIM}=\begin{pmatrix}
        {h}_0 & V_1 & \dots &  V_N\\
         \bar{V}^\dagger_1 & \Omega_1 I_1 & 0 & 0 \\
        \vdots & 0 & \ddots & 0\\
        \bar{V}^\dagger_N & 0 & \dots & \Omega_N I_N
    \end{pmatrix}.
    \label{eq:HAIM}
\end{equation}
As shown in Ref.~\cite{chiarotti_energies_2024}, the (generally complex) eigenvalues of the (possibly non-Hermitian) AIM Hamiltonian are the poles of the Green's function in Eq.~\eqref{eq:dyson_eq}, with a self-energy from Eq.~\eqref{eq:SOP_SE}, and the Dyson orbitals correspond to the projections of the AIM eigenvectors onto the Hilbert space of $h_0$.
Specifically, the Green's function of Eq.~\eqref{eq:dyson_eq} can be written as:
\begin{equation}
    G(\omega) = \sum_s \frac{P_{h_0} \ket{\psi^r_s} \bra{\psi^l_s} P_{h_0}}{\omega - z_s},
    \label{eq:SOP_G}
\end{equation}
where $h_0$ projects away the DOF of the baths, and $\ket{\psi^r_s}$ ($\bra{\psi^l_s}$) are the right (left) eigenvector of $H_\text{AIM}$ with eigenvalue $z_s$. 

The dimension of the fictitious bath, $\text{dim}[I_i]$ in the effective Hamiltonian, contributing to the number of poles/satellites of $G$ in Eq.~\eqref{eq:SOP_G}, is determined by the ranks of the residues.
Thus, the dimension of the AIM system follows: $\text{dim}[H_\text{AIM}] = \text{dim}[h_0] + \sum_m{\text{rank}[\Gamma_m]}$.
In the case of a localized self-energy, such as for the dynamical Hubbard functional, the rank of $\Gamma_i$ is bounded by the dimension of the localized manifold $\{\ket{\phi^{I}_{m}}\}_{I,m}$ from Eq.~\eqref{eq:unfolding_SE}, further reducing the computational cost compared to a fully non-local correction like $GW$\footnote{This can be seen from the singular values of the $k$-dependent dynamical-Hubbard self-energy residues, that, by construction, match those of the local self-energy.}.
For more details on the methodology, mathematical proofs, and additional references, see Refs.~\cite{chiarotti_energies_2024,chiarotti_spectral_2023}. 

Here, we note that similar methods were used in the context of DMFT and the Hubbard model to efficiently invert the Dyson equation~\cite{savrasov_many-body_2006,kotliar_electronic_2006}.
In the context of ab initio methods, similar SOP-based approaches that avoid explicit frequency representations of propagators have been employed in Refs.~\cite{savrasov_many-body_2006,bintrim_full-frequency_2021,backhouse_scalable_2021,bintrim_full-frequency_2022,scott_moment-conserving_2023,gao_efficient_2024}.
In particular, Refs.~\cite{leon_frequency_2021,leon_efficient_2023,leon_spectral_2025} exploit the SOP representations of the screened Coulomb interaction to go beyond the plasmon pole approximation also providing for a $G_0W_0$ self-energy on SOP.

\subsection{Charge self-consistent calculations} 
\label{sec:method_cscf}
In this Section, we develop first the simpler framework for charge self-consistent calculations with the dynamical Hubbard functional, crucially leveraging the AIM-SOP method; this will be extended to  self-consistency in the Green's function in the next Section.
In the charge self-consistent implementation of the dynamical Hubbard functional, we iteratively solve the Dyson equation (we omit here and henceforth the spin index):
\begin{equation}
    G(\omega)=
    \Big[\omega-h_\text{KS}[\rho]-\Sigma_\text{dynH}[\mathbf{G}_\text{KS}](\omega)\Big]^{-1},
    \label{eq:dyson_charge_scf}
\end{equation}
with the charge density $\rho$ and $\mathbf{G}_\text{KS}$ given by:
\begin{equation}
    \begin{split}
        &\rho(\mathbf{r})=\frac{1}{2} \int \frac{d\omega}{2\pi i} e^{i\omega0^+}G(\mathbf{r},\mathbf{r},\omega),\\
        &\mathbf{G}_\text{KS}(\omega)=\mathbf{P}_{H}G_\text{KS}[\rho](\omega)\mathbf{P}_{H},
    \end{split}
\end{equation}
where $h_\text{KS}$ is the Kohn–Sham (KS) Hamiltonian.

We use the AIM-SOP method described in Sec.~\ref{sec:AIM-SOP} to obtain the SOP representation of $G$ from the SOP of $\Sigma(\omega)$ given by Eq.~\eqref{eq:dynHUB_SE}~\footnote{The SOP for the self-energy is obtained from the SOP of $\mathbf{G}_\text{KS}(\omega)$ and $U(\omega)$ using the Cauchy residue theorem for the convolution of two SOP-propagators~\cite{chiarotti_unified_2022}.}.
We also use the Cauchy residue theorem to compute the charge density from the SOP of the Green’s function---this corresponds to the zeroth occupied moment of $G$; see Ref.~\cite{chiarotti_unified_2022}.

The use of AIM-SOP is essential, as self-consistent calculations demand high numerical accuracy (on the order of $\sim 10^{-8}$ Ry for the total energy, for the systems studied here).
For example, even a small error in the charge density would significantly affect this precision:  
a 1\% error in the integrated density would lead to the failure of the self-consistent cycle, as the total energy depends approximately linearly on the number of particles.
In addition, as previously discussed, the fundamental advantage of this method over imaginary-axis formulations lies in its spectral resolution, which is typically difficult to obtain (and is avoided here) from the analytic continuation procedures.

As shown in Eq.~\eqref{eq:dyson_charge_scf}, in the charge self-consistent scheme, only the charge density is updated self-consistently, while $\mathbf{G}$ is constrained to be $\mathbf{G}_\text{KS}[\rho]$, with the density of the current iteration.  
Explicitly, at each iteration, given a density $\rho(\mathbf{r})$, the poles and residues of the KS local Green’s function $\mathbf{G}_\text{KS}$ are:
\begin{equation}
    \begin{split}
        &z_{s}=\epsilon^\text{KS}_{\mathbf{k}n}+i\eta\text{sign}(\epsilon^\text{KS}_{\mathbf{k}n}-\mu^\text{KS}),\\
        &\mathbf{A}_s=\mathbf{P}_H\ket{\psi_{\mathbf{k}n}^\text{KS}}\bra{\psi_{\mathbf{k}n}^\text{KS}}\mathbf{P}_H \ .
    \end{split}
\end{equation}
Here, $\epsilon^\text{KS}_{\mathbf{k}n}$ and $\ket{\psi_{\mathbf{k}n}^\text{KS}}$ are the eigenvalues and eigenvectors of $h_\text{KS}[\rho]$, respectively.  
The parameter $\eta$ is a numerically small value (typically $\sim 10^{-9}$ eV) used to preserve the time-ordering of the propagator, and $\mu^\text{KS}$ is the chemical potential of the KS calculation.
Following our one-shot implementation~\cite{chiarotti_energies_2024}, we apply a smearing to $\mathbf{G}_\text{KS}$, which is detailed in Appendix~\ref{sec:smearingSOP}.

The energy functional minimized by the charge self-consistent scheme is:
\begin{equation}
     E_\text{dynH}^\text{c-scf}[G_\text{KS}] = E_\text{DFT}[\rho] + \Phi_{\text{dynH}}[\mathbf{G}_\text{KS}].
     \label{eq:ene_cscf}
\end{equation}
The minimization is stopped according to an energy convergence criterion:
\begin{equation}
     \Delta E_\text{dynH}^\text{c-scf}[G_\text{KS}] \le \text{thr},
     \label{eq:cscf_thr}
\end{equation}
where $\Delta$ denotes the energy difference between two successive iterations, and the threshold is typically set to $\lesssim 10^{-9}$ Ry (see also Sec.~\ref{sec:numerical_details}).

We adopt a linear mixing scheme to update the charge density: at iteration $n$, the mixed density is given by $\rho^\text{mix}_n(\mathbf{r})= (1-\beta)\rho_{n-1}(\mathbf{r}) + \beta\rho_{n}(\mathbf{r})$.
Additionally, we employ a pole condensation procedure on the local Green's function $\mathbf{G}$ at each iteration.  
This procedure was introduced in our Ref.~\cite{quinzi_broken_2025} and is described in Appendix~\ref{sec:condensationSOP} for completeness.
Within each iteration, we also perform a self-consistent search for the chemical potential to conserve the number of particles in the solid, $N$.  
This procedure is also described in Appendix~\ref{sec:chem_pot}.

\subsection{Fully self-consistent calculations}
\label{sec:method_fscf}
In this Section, we describe the framework for fully self-consistent calculations with the dynamical Hubbard functional, which constitute the main result of this paper.  
Leveraging the AIM-SOP method, the stationarization of a spectral energy functional is achieved with high numerical precision, while also guaranteeing (real-axis) spectral accuracy for dynamical quantities.  
This represents an important result for Green's function-based functional formulations, ensuring a numerical precision comparable to that of DFT, but now for a dynamical first-principles framework.
  
The fully self-consistent implementation of the dynamical Hubbard functional aims to stationarize the spectral (Klein) functional of Eq.~\eqref{eq:klein_locGW}, $E[G]$.  
The key difference with respect to the charge self-consistent scheme of Sec.~\ref{sec:method_cscf} is that the functional does not reduce to the form of Eq.~\eqref{eq:ene_cscf}, and all terms must be accounted for an accurate implementation.

We stationarize the functional with respect to the Green's function $G$ by iteratively solving the Dyson equation:
\begin{equation}
     G(\omega)= \Big[\omega-h_\text{KS}[\rho]-\Sigma_\text{dynH}[\mathbf{G}](\omega)\Big]^{-1},
\end{equation}
where the charge density and local Green’s function are updated at each iteration as:
\begin{equation}
    \begin{split}
        &\rho(\mathbf{r})= \int \frac{d\omega}{2\pi i} e^{i\omega0^+}G(\mathbf{r},\mathbf{r},\omega), \\
        &\mathbf{G}(\omega) = \mathbf{P}_H G(\omega) \mathbf{P}_H.
    \end{split}
    \label{eq:G_2_rho_Gbold}
\end{equation}
Here, $h_\text{KS}[\rho] = T + v_\text{ext} + v_{Hxc}[\rho]$ is the KS Hamiltonian evaluated at the density from Eqs.~\eqref{eq:G_2_rho_Gbold}, and $\Sigma_\text{dynH}[\mathbf{G}]$ is the dynamical Hubbard self-energy from Eq.~\eqref{eq:dynHUB_SE}.

As in the charge self-consistent case, the use of AIM-SOP to solve the Dyson equation is crucial.  
The same considerations on the numerical accuracy needed apply, with the added complexity that the SOP for the self-energy must now be constructed from Eq.~\eqref{eq:dynHUB_SE} using the SOP representation of $\mathbf{G}(\omega)$.  
In addition to the accurate charge density (computed via the Cauchy residue theorem as discussed in Sec.~\ref{sec:method_cscf}), the SOP for the localized Green's function $\mathbf{G}(\omega)$ is directly obtained from the SOP of $G(\omega)$ via:
$\mathbf{G}(\omega) = \sum_s \frac{\mathbf{A}_s}{\omega - z_s}, \quad \text{with } \mathbf{A}_s = \mathbf{P}_H A_s \mathbf{P}_H$.

The procedure for evaluating all terms in the energy functional is described in Appendix~\ref{sec:KleinSOP}.  
At each iteration $n$, we employ a linear mixing on the Green’s function, $G^\text{mix}_n(\omega) = (1-\beta) G_{n-1}(\omega) + \beta G_n(\omega)$.
As in the charge self-consistent case, pole condensation is applied to the local Green’s function $\mathbf{G}$ at each iteration.  
The chemical potential is updated to ensure particle-number conservation, following the same procedure as described for the charge self-consistent scheme in Appendix~\ref{sec:chem_pot}.

For initialization, we randomize the charge density over the atoms, following standard practice in plane-wave DFT codes such as in the PWscf \QE\ distribution.  
We then compute $v_{Hxc}[\rho]$ and $h_\text{KS}[\rho]$, evaluate $\Sigma_\text{dynH}$ using Eq.~\eqref{eq:dynHUB_SE} in SOP with the initial $G_\text{KS}$, and solve the Dyson equation~\eqref{eq:dyson_charge_scf} using AIM.  
From the second iteration onward, we obtain the charge density $\rho$ and the projected Green’s function $\mathbf{G}$ from Eqs.~\eqref{eq:G_2_rho_Gbold}, calculate the corresponding $v_{Hxc}[\rho]$ and $\Sigma_\text{dynH}[\mathbf{G}]$, evaluate the total dynamical Hubbard energy via Eq.~\eqref{eq:klein_locGW}, and again solve the Dyson equation using AIM.

As in the charge self-consistent case, we stop the self-consistent cycle once the energy variation satisfies:
\begin{equation}
     \Delta E_\text{dynH}^\text{scf}[G] \le \text{thr},
     \label{eq:scf_thr}
\end{equation}
The numerical details on the convergence threshold, condensation factor, Hubbard manifold, pseudopotentials, and other parameters are provided in Sec.~\ref{sec:numerical_details}.

\section{Results}

\begin{figure}
    \centering
    \includegraphics[width=\columnwidth]{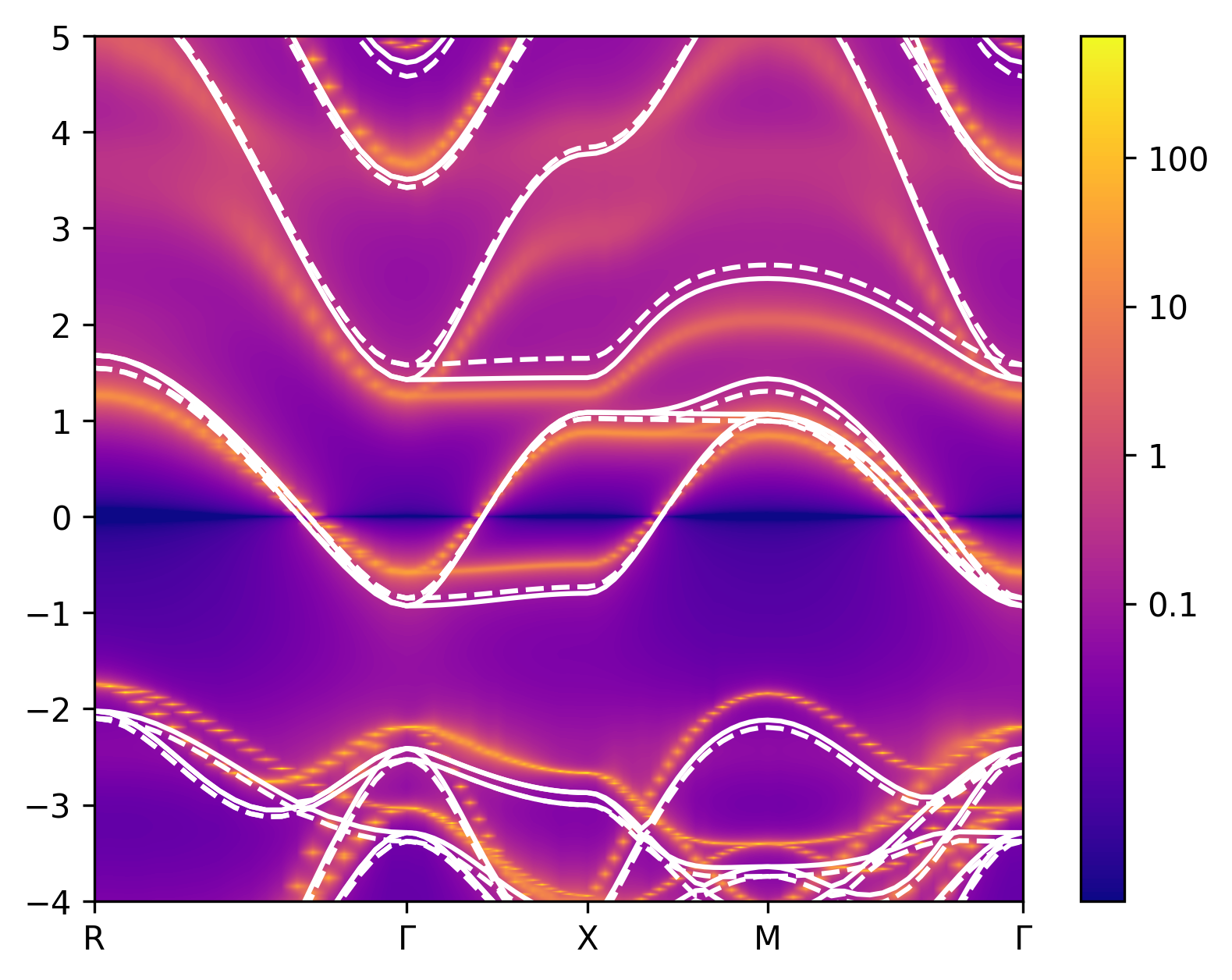}
    \caption{Fully self-consistent spectral function from the dynamical Hubbard functional of SrVO$_3$ (color plot, this work) compared to PBEsol (solid white line) and PBEsol + U (dashed white line). The chemical potential is aligned to $0$ in each case. The color map is logarithmic.}
    \label{fig:SF_SrVO3}
\end{figure}

\begin{figure}
 \centering
    \includegraphics[width=\columnwidth]{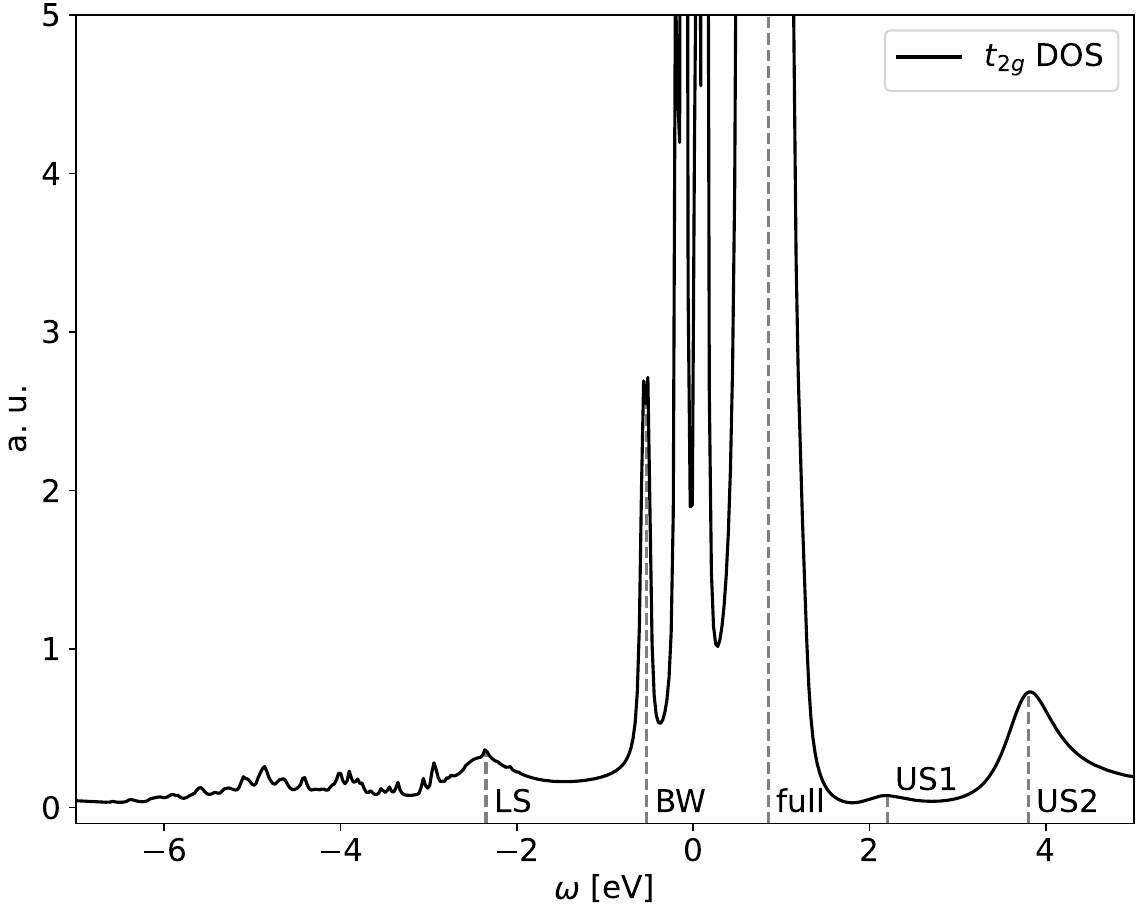}
    \caption{Fully self-consistent density of states from the dynamical Hubbard functional of the $t_{2g}$ bands of SrVO$_3$ (this work). The lower satellite (LS), quasi-particle occupied bandwidth (BW), full bandwidth (full), and upper satellites (US1 and US2) are marked with grey dashed lines~\cite{notes_for_DOS}. The chemical potential is shifted to $0$.}
    \label{fig:DOS_SrVO3}
\end{figure}

\begin{table}
    \begin{ruledtabular}
        \begin{tabular}{lcccc}
             {\bf Method}       & {\bf BW} & ${\mathbf{m}^*}/{\mathbf{m_\text{\bf PBEsol}}}$    & {\bf LS} & {\bf US} \\
            \hline
                PBEsol     &      $1.0$   &     $1$     &            \\
                PBEsol + U &  $0.92$   &     $1.1$  &           &    \\
            \hline
                GW+EDMFT~\cite{boehnke_when_2016}&    $0.5$   &    $2$        &  $-1.7$     &    $2.8$     \\
            \hline
            PBEsol+dynH  &   $0.5$   &    $2$   & $-2.5 \pm 1$ &  $\begin{matrix} 2.6 \\ 3.5 \end{matrix} \pm 1$  \\
            scf-PBEsol+dynH  &   $0.5$   &    $1.4$   & $-2.35 \pm 1$ &  $\begin{matrix} 2.2 \\ 3.8\end{matrix} \pm 1$  \\
            \hline
                exp.~\cite{yoshida_direct_2005} &   $0.7$    &    $1.8$   &  $-1.5$                \\
                exp.~\cite{takizawa_coherent_2009} &   $0.44$   &    $2$    &  $-1.5$             \\
        \end{tabular}
    \end{ruledtabular}
    \caption{Occupied bandwidth (BW) 
    %\AP{\footnote{The occupied bandwidth is found by taking the energy difference between the fermi energy and the highest occupied quasiparticle peak in the DOS}} 
    of the $t_{2g}$ bands, the mass enhancement factor (${{m}^*}/{{m}_\text{{PBEsol}}}$), and energy of the lower (LS) and upper satellites (US1 and US2) from experiments and different theoretical frameworks ---with two numbers indicating two different satellites.
    One-shot values from the dynamical Hubbard functional (PBEsol+dynH) are from Ref.~\cite{chiarotti_energies_2024}.
    Fully self-consistent dynamical Hubbard functional results (scf-PBEsol+dynH, this work) are estimated from the spectral function in Fig.~\ref{fig:SF_SrVO3} and the
    t$_{2g}$ DOS in Fig.~\ref{fig:DOS_SrVO3}. 
    As for the one-shot case in Ref.~\cite{chiarotti_energies_2024}, ${{m}
    ^*}/{{m}_\text{{PBEsol}}}$ is estimated using the ratio of the full bandwidths. All energies are in 
    eV.}
    \label{tab:spectrum_SrVO3}
\end{table}

\begin{figure}
    \centering
    \includegraphics[width=\columnwidth]{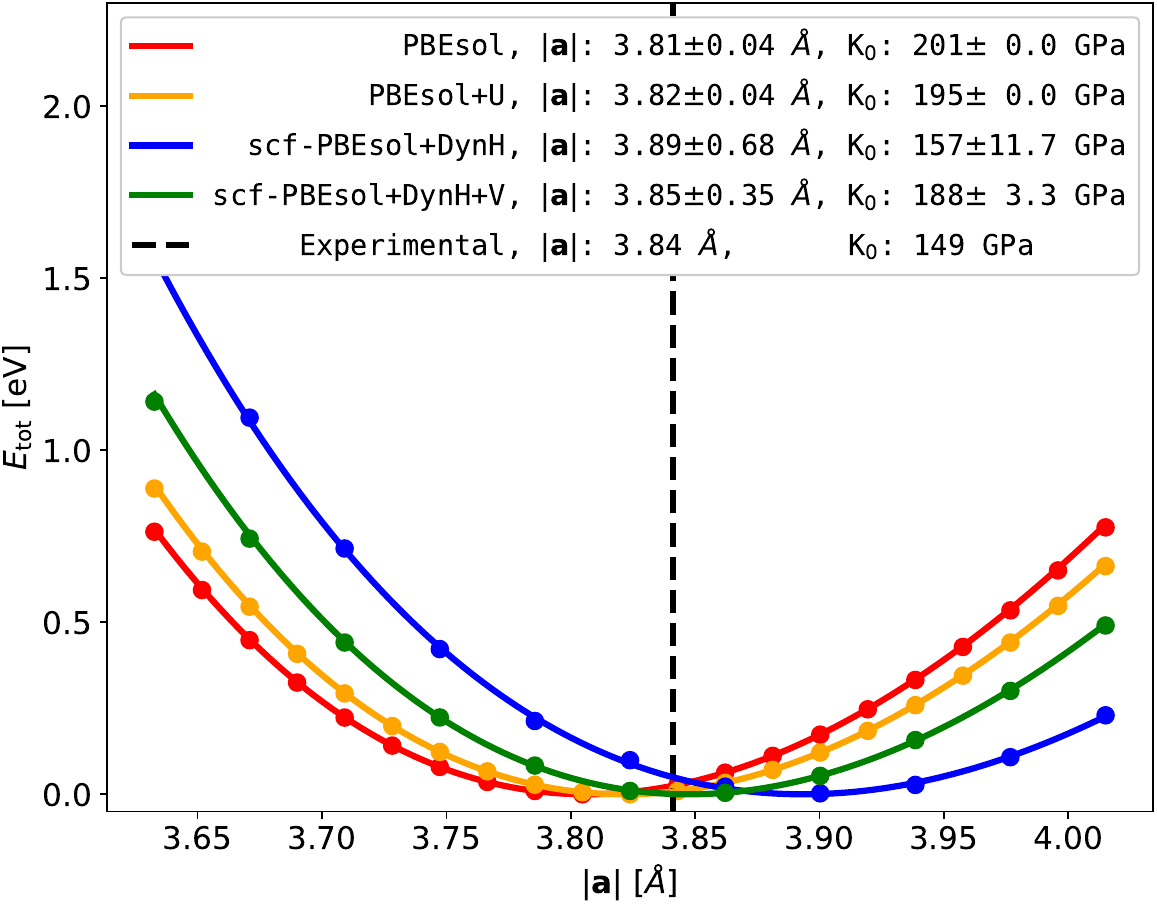}
    \caption{Equation of state for SrVO$_3$ calculated using PBEsol in red, PBEsol + U in orange and the fully self-consistent dynamical Hubbard method (scf-PBEsol+dynH) from this work in blue.
    In green we plot the results of scf-PBEsol+dynH with an additional $+V$ correction to account for intersite interactions.
    Data (dots) are fitted considering a Birch-Murnaghan dependence (solid line). 
    The values and errors from the fit are displayed in the legend. 
    For reference, the experimental lattice parameter is marked in dashed black, with the bulk modulus also reported above.
    We omit the DFT+DMFT predictions~\cite{haule_free_2015} as they are very similar to PBEsol; for a detailed comparison between the theories refer to the main text.
    Experimental values are taken from~\cite{maekawa_physical_2006}.}
    \label{fig:bulk_mod}
\end{figure}

Leveraging the dynamical Hubbard functional and the algorithmic-inversion method on sum over poles, we perform both charge and fully self-consistent calculations on SrVO$_3$ as a paradigmatic test case for the theory.  
To account for intersite interactions, we also include a static $+V$ correction within the fully self-consistent framework~\footnote{Please note that here with $+V$ correction we mean we perform a DFT+V (static) calculation with the value of $V$ from linear response~\cite{timrov_hubbard_2018,timrov_accurate_2022}.}.  
These results extend and complete the one-shot calculations presented in Ref.~\cite{chiarotti_energies_2024}.

In Fig.~\ref{fig:SF_SrVO3}, we show the self-consistent spectral function of SrVO$_3$, obtained using PBEsol, PBEsol+$U$, and the fully self-consistent dynamical Hubbard framework (scf-DFT+dynH), marked as ``scf-PBEsol+dynH'' in the Figure.  
The latter is qualitatively similar to the one-shot DFT+dynH result~\cite{chiarotti_energies_2024}, as well as to state-of-the art GW+EDMFT calculations~\cite{boehnke_when_2016} and experimental measurements~\cite{yoshida_direct_2005,takizawa_coherent_2009}.  
A detailed comparison between theory and experiments is in Table~\ref{tab:spectrum_SrVO3}.
The reduction of the bandwidth to nearly half compared to PBEsol, along with a quasiparticle mass enhancement factor of approximately $1.4$, highlights the improvement in quasiparticle properties relative to (PBEsol) DFT~\cite{notes_for_DOS}.  
The reduction in mass enhancement from $2$ to $1.4$ also reveals slightly asymmetric t$_{2g}$ bands with respect to the Fermi energy, a feature also observed in experiments~\cite{takizawa_coherent_2009}.  
Looking at the t$_{2g}$ density of states (DOS) in Fig.~\ref{fig:DOS_SrVO3}, the position of the lower satellite (LS) lies slightly below the experimental measurement, albeit with sufficiently broad features, similar to the one-shot case.  
The first and second upper satellites (US1 and US2) also appear in the same positions as in the one-shot calculations and remain compatible with GW+DMFT results.

Moving to the thermodynamic properties in Fig.~\ref{fig:bulk_mod}, all methods---PBEsol, PBEsol+$U$, and DFT+DMFT from Ref.~\cite{haule_free_2015}---yield reasonably accurate predictions for the equilibrium lattice parameter, although they tend to slightly overestimate the stiffness of the material.  
The bulk modulus predicted by the fully self-consistent dynamical Hubbard framework---both without (termed scf-PBEsol+dynH in Fig.~\ref{fig:bulk_mod}) and with the $+V$ intersite term (termed scf-PBEsol+V+dynH in Fig.~\ref{fig:bulk_mod})---is reduced, confirming the one-shot prediction.  
Also, an almost perfect agreement with the experimental equilibrium lattice parameter from Ref.~\cite{maekawa_physical_2006} is obtained using the fully self-consistent DFT$+V$+dynH scheme~\footnote{As mentioned in the numerical details, the experimental lattice parameter is extrapolated to zero temperature using the thermal expansion coefficient reported in the same reference.}.  
Here, scf-DFT+dynH overestimates the lattice parameter by $1.3\%$ (compared to a $\sim 0.9\%$ overestimate by PBEsol) and the volume by approximately $4\%$.
Beyond the excellent agreement with experiments, the comparison between DFT+$U$ and the dynamical Hubbard results shows that the renormalization of thermodynamic, frequency-integrated quantities, can arise from dynamical effects.

The spectral properties predicted by the charge self-consistent dynamical Hubbard method are identical from those in Fig.~\ref{fig:SF_SrVO3} and thus omitted here.  
The predicted lattice parameter is $3.91$\AA\ and the bulk modulus is $159$~GPa.  
These findings demonstrate that charge self consistency is sufficient to reproduce both the experimental results and the fully self-consistent predictions for the spectral properties of SrVO$_3$, while being cheaper and less computationally challenging to implement.  
However, we note that this observation may not hold in general---particularly for more complex materials such as Sr$_2$RuO$_4$.

\section{Numerical details}
\label{sec:numerical_details}
All DFT and DFT+U calculations are performed using the PWscf code of the \textsc{Quantum ESPRESSO} 
distribution~\cite{giannozzi_quantum_2009}.
Calculations with the dynamical Hubbard functional are obtained using the (in house) dynamical Hubbard code.
To calculate $v_\text{Hxc}[\rho]$ at each self-consistent iteration, for both DFT+dynH CSCF and SCF self-consistent schemes, PWscf is used.
The localized screened-potential $U(\omega)$ is calculated with RESPACK~\cite{nakamura_respack_2021} using maximally localized Wannier functions.
The pseudopotentials used for all the calculations are optimized norm-conserving pseudopotentials~\cite{hamann_optimized_2013} (PBEsol standard-precision, nc-sr-04) from the {\sc Pseudo Dojo} library~\cite{van_setten_pseudodojo_2018}.

As done in the one-shot case~\cite{chiarotti_energies_2024}, the spectral results in Figs.~\ref{fig:SF_SrVO3} and~\ref{fig:DOS_SrVO3} are obtained using a simple-cubic cell, with a lattice parameter of $a=3.824$~\AA. 
This is the extrapolated value at the zero-temperature limit of the experimental value $a=3.841$ \AA \ from Refs.~\cite{maekawa_physical_2006} and~\cite{lan_structure_2003}, 
considering an average linear thermal expansion coefficient $\alpha_l = 1.45 \times 10^{-5} K^{-1}$ from Ref.~\cite{maekawa_physical_2006}.
The $k$ mesh for these calculation is $6\times6\times6$ and symmetries are exploited to reduce the number of $k$ points.
We use a Marzari-Vanderbilt~\cite{marzari_thermal_1999} smearing of $\sim 0.27$ eV.
With these parameters we assure a convergence of $\sim 0.1$ \AA$^3$ \ and $\sim 0.1$ GPa on the equilibrium lattice parameter and the bulk modulus using PBEsol.
For charge self-consistent calculations we broaden $G_\text{KS}$ with a numerically vanishing amount of $i \eta_\pm=\pm i \ 10^{-9}$ eV, to distinguish whether the pole is occupied ($+$) or empty ($-$) below and above the chemical potential, respectively.
We use $2$ eV threshold for the pole condensation; for further details on this procedure we refer to Appendix~\ref{sec:condensationSOP}. 
Calculations with the condensation threshold reduced to $1$ eV give a difference of 0.2 \AA$^3$ \ for equilibrium volume, and $10$ GPa for the DFT+dynH SCF bulk modulus; both are within the error bar from the fit.
The value of the self-consistent thresholds of Eqs.~\eqref{eq:cscf_thr} and~\eqref{eq:scf_thr} is $10^{-7}$~Ry---with a possible and continuous lowering if required.
To plot the band structure in Fig.~\ref{fig:SF_SrVO3} we have added a broadening of all the poles of the Green's function (noticeable only for quasiparticle bands) by $10^{-3}$~eV to help the visualization.
\section{Conclusions}
\label{sec:conclusions}
In this paper, we present a self-consistent framework based on a (semi)local and dynamical functional to treat correlated materials from first principles.  
Specifically, we augment the exchange-correlation part of DFT with a dynamical Hubbard term (DFT+dynH), generalizing DFT+$U$ to include a frequency-dependent screening $U(\omega)$~\cite{chiarotti_energies_2024}.  
This leads to a spectral energy functional of the Green's function, which is variational and requires the self-consistent solution of the Dyson equation to be stationary. 

Crucially, self consistency is made possible by the algorithmic-inversion method based on a sum-over-poles representation (AIM-SOP), ensuring the needed numerical accuracy for the calculations while preserving (real-axis) spectral resolution for dynamical properties.  
As a result, the present approach avoids the issues associated with analytic continuation---commonly encountered in imaginary-axis formulations~\cite{kutepov_one-electron_2017,fei_nevanlinna_2021,iskakov_greenweakcoupling_2025}---and achieves numerical accuracy comparable to that of DFT (on the order of $\sim 10^{-9}$ Ry for the total energy), for both spectral and thermodynamic quantities.  
In this work, we develop both charge self-consistent and fully self-consistent schemes, with the latter representing the primary result of the paper.  
Using this method, we self-consistently compute the spectral and thermodynamic properties of SrVO$_3$.  

As shown in Refs.~\cite{chiarotti_energies_2024,caserta_dynamical_2025}, already the one-shot spectral results from DFT+dynH exhibit remarkable agreement with both experimental data and state-of-the-art $GW$+EDMFT calculations for SrVO$_3$—a paradigmatic metallic perovskite—as well as for four Mott-Hubbard/charge-transfer insulators: MnO, FeO, CoO, and NiO.  
While for these systems one-shot results suffice for spectral properties, accurate thermodynamic predictions are significantly more delicate and require a fully self-consistent formulation of the theory.

In Figs.~\ref{fig:SF_SrVO3} and \ref{fig:DOS_SrVO3}, and in Table~\ref{tab:spectrum_SrVO3}, we show that the fully self-consistent dynamical Hubbard calculations calculations do not qualitatively alter the spectral function of SrVO$_3$ compared to the one-shot results, which already align with $GW$+EDMFT and experimental data~\cite{boehnke_when_2016,yoshida_direct_2005,takizawa_coherent_2009}---charge self-consistency is omitted giving the same results.  
Turning to thermodynamic predictions, the equilibrium lattice parameter shown in Fig.~\ref{fig:bulk_mod} is lowered with respect to the one-shot prediction, becoming more compatible with experiments.
The overestimation present in the one-shot case by $2.1\%$ in the lattice parameter is lowered to $1.3\%$ self-consistent approach (scf-DFT+dynH), and to below $0.05\%$ when a (static) $+V$ correction is included to account for intersite interactions.
Furthermore, we confirm the reduction in the bulk modulus observed in the one-shot calculations, with the self-consistent values ranging from $157 \pm 11.7$ GPa to $188 \pm 3.3$ GPa---the latter being with the $+V$ correction.  
In addition to the excellent agreement with experiment, comparison with DFT+$U$ shows that the renormalization of frequency-integrated thermodynamic quantities can originate from dynamical effects.

Beside the specific case study, the comparable numerical accuracy of the present method with DFT for both spectral and thermodynamic properties---together with its consistent treatment of dynamical effects such as band renormalization, mass enhancement, and spectral-weight transfer---marks a fundamental step in the investigation of correlated materials fully from first principles, using a spectral functional theory that goes beyond DFT.

%========================
\section{Acknowledgments}
\label{sec:Acknowledgments}
%========================
This work was supported by the Swiss National Science Foundation (SNSF) through grant No. 200020-213082 (T.C.,M.Q.,A.P.,N.M.) and NCCR MARVEL (N.M.), a National Centre of Competence in Research through grant No.~205602.
M.C. was funded by a Bosch Research Foundation in the German Stifterverband grant, and A.F. by the EU Commission for the MaX Centre of Excellence on ‘Materials Design at the eXascale’ under grant No.~101093374.
The authors wish to thank Massimo Capone for many fruitful discussions. 

\appendix
\section{Evaluation of spectral (Klein) energy functionals on SOP}
\label{sec:KleinSOP}
In this Appendix we discuss the evaluation of the total energy functional of Eq.~\eqref{eq:klein_locGW} within the SOP formalism.
%
%\begin{align}
%    \label{eq:phi_eval}
%    \Phi[\mathbf{G}] &= \frac{1}{2}\int \frac{d\omega}{2\pi i}e^{i\omega 0^+} \mathbf{G}(\omega)\mathbf{\Sigma}_\text{dynH}(\omega)\\
%    &=\frac{1}{2}\left(\sum_s^{occ} \mathbf{A}^s\mathbf{\Sigma}_\text{dynH}(z_s) + \sum_m^{occ}\mathbf{G}(\Omega_m)\mathbf{\Gamma}^m\right)
%\end{align},
%
Specifically, we exploit the meromorphic structure of the dynamical propagators together with Cauchy residue theorem to give an explicit form of the terms in Eq.~\eqref{eq:klein_locGW}.

We start by considering $\Trw[I - G_0^{-1}G]$:
\begin{align}
    \label{eq:TrIGoG_eval}
    \nonumber
    \Trw[I - G_0^{-1}G]& =
    %\Trw\left[I - \left(\omega I-h_0\right)G(\omega)\right]\\  
%    \nonumber
    \Trw\left[I - \sum_s \frac{\omega A_s }{\omega -z_s} + h_0G(\omega)\right]\\
%    \nonumber
%    &= \Trw\left[I - \sum_sA^s -\sum_s \frac{z_s A^s}{\omega -z_s} + h_0 G(\omega)\right]\\
    &= -\sum_s^{\text{Im}[z_s]>0}z_s\text{Tr}\left[A_s\right] + \text{Tr}\left[h_0 \gamma\right],
\end{align}
%\begin{align}
%    \label{eq:TrIGoG_eval}
%    \Trw[I - G_0^{-1}G]=-
%\end{align}
where in the last line we used the definition of the density matrix $\gamma(\mathbf{r},\mathbf{r'}) = \int \frac{d\omega}{2\pi i}e^{i\omega 0^+}G(\mathbf{r},\mathbf{r'},\omega)$, and the completeness over the Dyson orbitals $\sum_s A_s = I$---exactly enforced within the AIM construction, see Ref.~\cite{chiarotti_unified_2022}.
The first term of Eq.~\eqref{eq:TrIGoG_eval} can be readily obtained from the poles and residues of the Green's function, while the second involves the noninteracting Hamiltonian $h_0 = T + v_{ext}$.
This can be evaluated by noting that $\text{Tr}\left[h_0 \gamma\right]=\text{Tr}[h_\text{KS}\,\gamma]-\text{Tr}[v_\text{Hxc}[\rho]\, \rho]$.

The $\Trw\text{ln}\left[G_0^{-1}G\right]$ can be computed analytically following our result of Ref.~\cite{ferretti_greens_2024}:
\begin{align}
    \label{eq:Trln_eval}
    \nonumber
    \Trw&\text{ln}[G_0^{-1}G] =\\ &\sum_s^\text{occ} z_s \text{rank}[A_s] - \sum_m^\text{occ}\Omega_m \text{rank}[\Gamma_m] - \sum_n^\text{occ}\varepsilon_n^0 \text{rank}[A^0_n], 
\end{align}
where $\Omega_m$ and $\Gamma_m$ are respectively the poles and residues of the self-energy operator, here defined as $\Sigma~=~G_0^{-1}~-~G^{-1}$, and $\varepsilon^0_n$ and $A^0_n$ are the poles and residues of $G_0$.
At variance with Eq.~\eqref{eq:TrIGoG_eval}, the poles in Eq.~\eqref{eq:Trln_eval} are now weighted with the ranks of their residues to account for the multiplicity of the corresponding eigenvalue within the algorithmic-inversion matrix.
For a noninteracting Green's function, in the absence of smearing, the rank of each residue is equal to its trace, and the last contribution of Eq.~\eqref{eq:Trln_eval} cancels with the term $\Trw[h_0 G_0]=\sum_n^\text{occ}\varepsilon_n^0\text{Tr}[A_n^0]$, avoiding the complication of treating it explicitly.
%\MQ{In the case of smearing, the traces of the residues of $G_0$ are equal to their ranks times the occupation function $f(\varepsilon_n^0)$. However, since $G_0 = (\omega I - T -v_\text{ext})^{-1}$ is not updated in the self-consistent procedure, possible discrepancies between the two contributions would only give a constant shift in the total energy.}
%\MQnote{If smearing is included the $\Trw\text{ln}[G_0^{-1}G]$ term needs to be generalized, a possibility would be to multiply the ranks of the residues by the occupation function, but this was numerically unstable for the evaluation of the bulk modulus with Marzari-Vanderbilt smearing. It only worked well for noninteracting $G$.
%Other possible route, one should re-evaluate the self-energy as $\Tilde{\Sigma}=G_{0}^{-1}-G_{smear}^{-1}$.}
The numerically stable evaluation of Eq.~\eqref{eq:Trln_eval} is made possible by computing this term just after the solution of the Dyson equation via the AIM-SOP as detailed in Sec.~\ref{sec:AIM-SOP}, before smearing and condensation of the Green's function poles are applied (see Appendix~\ref{sec:condensationSOP} and~\ref{sec:smearingSOP}).

The final expression for the total energy functional is then given by
%\begin{align}
%    \label{eq:etot_eval}
%    E =& E_\text{tot}^\text{DFT}[\rho] +  \sum_{n,\mathbf{k}}\bra{\psi_{n\mathbf{k}}}(\gamma - \gamma^\text{KS})\ket{\psi_{n\mathbf{k}}} +\Phi[\mathbf{G}]\\
%    \nonumber
%    &+ \sum_s^\text{occ} z_s \left(\text{rank}[A^s] - \text{Tr}[A^s]\right)  - \sum_m^\text{occ}\Omega_m\,\text{rank}[\Gamma^m].
%\end{align}
\begin{align}
    \label{eq:etot_eval}
    E_\text{dynH} =& E_\text{H}[\rho] + E_{xc}[\rho] + \Phi_\text{dynH}[\mathbf{G}] + \text{Tr}[h_0 \gamma] \\
    \nonumber
    &+ \sum_s^\text{occ}\,z_s\,\left(\text{rank}[A_s]-\text{Tr}[A_s]\right) - \sum_m^\text{occ}\,\Omega_m\,\text{rank}[\Gamma_m]
\end{align}
%We also stress that, while the self-energy poles and residues appear explicitly in Eq.~\eqref{eq:etot_eval}, the Klein functional is defined entirely in terms of the interacting and noninteracting Green's function, the self-energy being formally defined as $\Sigma=G_0^{-1}-G^{-1}$.
In passing we stress that the expression of Eq.~\eqref{eq:etot_eval} holds also for generalized KS-DFT functionals, e.g., depending on the localized occupation matrix---as in the case of DFT+V augmented by the dynamical Hubbard term.

Finally, the dynamical Hubbard $\Phi_\text{dynH}$ is expressed as:
\begin{align}
    \label{eq:phi_eval}
    \nonumber
    \Phi_\text{dynH}[\mathbf{G}] =& \frac{1}{2}\sum^{\text{Im}[z_{s_1}]>0}_{s_1} \sum_{t}^{\text{Im}[\theta_t]<0}\sum_{s_2}^{\text{Im}[z_{s_2}]<0} \frac{\text{Tr}[\mathbf{A}_{s_2} B_{t} \mathbf{A}_{s_2}]} {z_{s_1} - \theta_t - z_{s_2}}\\
    \nonumber
    +&\frac{1}{2}\sum^{\text{Im}[z_{s_1}]<0}_{s_1} \sum_{t}^{\text{Im}[\theta_t]>0}\sum_{s_2}^{\text{Im}[z_{s_2}]>0} \frac{\text{Tr}[\mathbf{A}_{s_2} B_{t} \mathbf{A}_{s_2}]}{z_{s_1} - \theta_t - z_{s_2}}\\
    +&\frac{1}{2}U_{\infty}\text{Tr}[\mathbf{\gamma}(\mathbf{I}-\mathbf{\gamma})]
\end{align}
where $\theta_t$ and ${B_t}$ are the scalar poles and residues of $U(\omega)$, $\mathbf{G}$ is the localized Green's function of Eq.~\eqref{eq:G_2_rho_Gbold}, with poles $z_s$ and residues $\mathbf{A}_s$, and $\mathbf{\gamma}$ is the localized occupation matrix obtained from $\mathbf{G}$.
Also, from the SOP expression above one easily deduce that the dynH functional reduces to the DFT+U functional in case of a constant screening $U(\omega)=U_\infty$.
Please note that while the $\Phi_\text{dynH}$ term in the spectral (Klein) functional is specific to this work (dynamical Hubbard), Eq.~\eqref{eq:etot_eval} is universal and can be used for other approximations of $\Phi_{xc}$.

\begin{figure}
    \centering
    \includegraphics[width=\columnwidth]{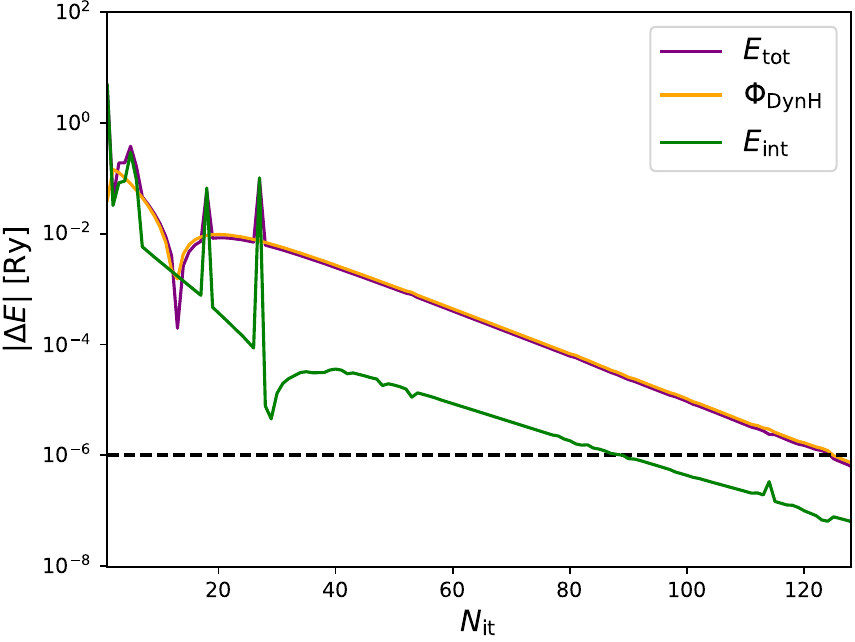}
    \caption{
    Convergence of the total energy computed with Eq.~\eqref{eq:etot_eval}.
    The contributions to the total energy functional are also separately reported, namely the dynamical Hubbard $\Phi_\text{DynH}$ term calculated with Eq.~\eqref{eq:phi_eval} and
    %the correction to the DFT band energy $E_\text{corr}$ calculated with Eq.~\eqref{eq:ecorr_eval} and
    $E_\text{int}$ computed from the second line of Eq.~\eqref{eq:etot_eval}.
    The energy differences between two successive iterations are reported in logarithmic scale and the threshold of $10^{-6}$ is reported with a horizontal dashed line as a guide to the eye. 
    The linear trend of the curves with respect to the number of iterations reveals the exponential convergence of the total energy.
    }
    \label{fig:Ediff_run}
\end{figure}

In Fig.~\ref{fig:Ediff_run} we show the convergence of the total energy with respect to the number of iterations in a full self-consistent calculation, with energy differences reported on a logarithmic scale.
In the plot, we report the value of the total energy computed from the dynamical Hubbard functional following Eq.~\eqref{eq:etot_eval}, together with the different contributions to the functional. 
In particular,
%the correction to the band energy $E_\text{corr}$ is calculated from Eq.~\eqref{eq:ecorr_eval} and in Fig.~\ref{fig:Ediff_run} 
we use $E_\text{int}$ as a shorthand for the second line of Eq.~\eqref{eq:etot_eval}, i.e., $E_\text{int}=\sum_{s}^\text{occ}z_s(\text{rank}[A^s]-\text{Tr}[A^s])-\sum_m^\text{occ}\Omega_m\text{rank}[\Gamma^m]$.
Please note that $E_\text{int}$ vanishes in the case of independent particle Green's functions.
The total energy is exponentially converging with respect to the number of iterations of the self-consistent loop.
For SrVO$_3$, all the different contributions share similar convergence rates, as observed from the slope of the curves, with the $\Phi_\text{dynH}$ term having the largest variations in absolute value. 
Spikes in the curves, particularly evident in the first steps of the self-consistent loop, are typically related to a significant variation of the total number of poles in the Green's function, which is mostly signaled by the $E_\text{int}$ term.

\section{Pole condensation}
\label{sec:condensationSOP}
In this Section we detail on the pole condensation procedure, which has the scope to reduce the number of poles in the self-energy, in turn controlling the dimension (together with the ranks of the residues) of the algorithmic-inversion matrix $H_\text{AIM}$, see Sec.~\ref{sec:AIM-SOP} for the discussion.
This procedure was introduced in Refs.~\cite{chiarotti_energies_2024,chiarotti_spectral_2023} and then further improved~\cite{quinzi_broken_2025}, also following~\cite{puig_von_friesen_kadanoff-baym_2010}.
Here, for dynamical Hubbard calculations, we condense the poles of the localized Green's function $\mathbf{G}$, which in turn reduces the number of poles in the self-energy in Eq.~\eqref{eq:dynHUB_SE}.
%In order to condense the poles, we adopt a strategy inspired by the weighted mean of $N$ random variables.
%Focusing on the $i$-th contribution to $\mathbf{G}(\omega)$ written as a SOP, thus on $\frac{\mathbf{A}_i}{\omega-z_i}$, the imaginary part can be thought as a (almost) 
%Lorentian distribution centered around the real part of the pole $\text{Re}[z_i]$ and with broadening is $\text{Im}[z_i]$\footnote{Note that since the amplitudes $\mathbf{A}_i$ are complex, this statement is not exact, but still provides for a motivation of the pole condensation described}.
%The core idea behind this pole condensation is that the overlap of two Lorentzians is greater than a threshold, the two poles can be regarded effectively as one.
% Let $\frac{\mathbf{A}_i}{\omega - z_i}$ be the $i$-th contribution to $\mathbf{G}(\omega)$ written as a SOP. 
The general scope of a pole condensation procedure is to substitute the SOP for $\mathbf{G}$ with another SOP having fewer poles.
Here, we do not only want to preserve the spectral resolution as much as possible, but also to conserve frequency-integrated quantities as well, which are crucial for numerical accuracy.
The employed procedure conserves the zeroth and first occupied moments of $\mathbf{G}$, that are related to the number of particles and energy of the system---see below for details.

Specifically, to condense the poles:
\begin{enumerate}
    \item We sequentially browse the ordered pole array (looking at their real parts).
    \item When the absolute difference between the real parts of two adjacent poles $z_s$ and $z_{s+1}$ is smaller than a threshold, we replace the couple with a new pole defined as:
    \begin{equation}
    \label{eq:cond_weight}
        z_s^\text{new} = \frac{w_s z_s + w_{s+1} z_{s+1}}{w_s+w_{s+1}},
    \end{equation}
    with $w_s = \abs{\text{Tr}[\mathbf{A}_s]}$ as weights.
    The amplitude $A_s^\text{new}$ of the new pole is given by the sum of the amplitudes of the previous poles.
    \item We move to the next adjacent poles and perform the same analysis.
    \item We repeat this procedure iteratively (from point 1), until the number of poles remain invariant, meaning that two adjacent new poles have to be larger than the given threshold.
\end{enumerate}
We perform this procedure separately for the four quadrants of the complex plane, separated  by the real axis and the position of the chemical potential.
We check the accuracy of the condensation procedure by converging the dynamical Hubbard energy from Eq.~\eqref{eq:dynH_functional} to meV accuracy~\footnote{Please note that for this work we converge total energy differences as we are interested in the equation of state, see Sec.~\ref{sec:numerical_details} for the exact value.}.
Importantly, the condensation algorithm above is particle conserving (zeroth-order occupied momentum~\cite{chiarotti_unified_2022}) and energy conserving (first-order occupied momentum~\cite{chiarotti_unified_2022}) in the sense of Galitskii-Migdal~\cite{galitskii_application_1958}.

To see this, we define the regularized $m$-th moment of the Green's function as~\cite{chiarotti_unified_2022}: 
\begin{align}
E_m[G] &= \oint_\mathcal{C} \frac{d z}{2\pi i} e^{i z 0^+} z^m \text{Tr}[G(z)]\\
%\nonumber
%&= \oint \frac{d z}{2\pi i}e^{i z 0^+}\text{Tr}\left[I + \sum_s \frac{z_s A_s}{z - z_s}\right]\\
\nonumber
&= \sum_s^{\text{Im}[z_s]>0} z_s^m \text{Tr}[A_s],
\end{align}
where the integration contour $\mathcal{C}$ is defined in the upper half of the complex plane and the evaluation of the integral can be performed analytically with Cauchy's residue theorem.
When condensing two poles $z_1$ and $z_2$ into a new pole $z_s$, the zeroth moment of the distribution is conserved by defining the new residue $A_s$ as the sum of the residues of the two poles:
\begin{equation}
    \label{eq:cond_zero_moment}
    \text{Tr}[A_s]=\text{Tr}[A_1]+\text{Tr}[A_2].
\end{equation}
The conservation of the first moment $E_1[G]$ gives instead the position of the new pole $z_1 \text{Tr}[A_1] + z_2 \text{Tr}[A_2] = z_s \text{Tr}[A_s]$:
\begin{align}
\label{eq:cond_first_moment}
    z_s =  \frac{z_1 \text{Tr}[A_1] + z_2 \text{Tr}[A_2]}{\text{Tr}[A_1]+\text{Tr}[A_2]},
\end{align}
where in the second line we used $A_s$ as defined in Eq.~\eqref{eq:cond_zero_moment}.
To ensure positive semi-definite weights in Eq.~\eqref{eq:cond_weight}, we take the absolute value of the trace of the residues appearing in Eq.~\eqref{eq:cond_first_moment}~\footnote{From the Lehmann representation, the Green's function is expected to have positive semi-definite residues. 
However, the positive semi-definiteness is not guaranteed in general in the case of complex valued poles, and thus we take the absolute value of the trace.}.
As can be seen, our condensation procedure respects both of the constraints, leaving the particle number and the Galitskii-Migdal energy conserved~\footnote{
The conservation of the Galitskii-Migdal energy is evident from the expression:
\begin{equation}
    E_\text{GM} = \frac{1}{2} \int \frac{d\omega}{2\pi i} e^{i0^+\omega} 
    \left[\omega+h_0\right] G(\omega) \, .
\end{equation}}.

%This condensation scheme, however, is not variational, as it does not preserve the functional.  
%Therefore, in principle, it could pose issues for the self-consistent procedure.  
%In practice, this was generally not a problem, as it did not affect the precision of the self-consistency conditions in Eqs.~\eqref{eq:cscf_thr} and~\eqref{eq:scf_thr}, with the rare exception of (easy to spot) oscillatory convergence behavior at high and specific values of the condensation threshold.

\section{Smearing propagators on SOP}
\label{sec:smearingSOP}
In this Section we describe the procedure to apply a smearing to propagators expanded on SOP.
Although general, in this work we use this technique to smear the Green's function $G(\omega)$ to be able to treat metallic systems, such as SrVO$_3$.
This technique was also used in Refs.~\cite{chiarotti_energies_2024,caserta_dynamical_2025}.
Let us consider the Green's function in the SOP form of Eq.~\eqref{eq:SOP_G}.  
Before smearing is applied, this corresponds to a time-ordered propagator with complex poles---with a non-vanishing imaginary parts in fully self-consistent calculations---$z_i$ above (below) the real axis for $\text{Re}z_i \le \mu$ ($\text{Re}z_i > \mu$), with $\mu$ the Fermi energy.
Following, e.g.~\cite{hellgren_discontinuities_2012}, in order to apply a smearing, we substitute each element in the SOP for $G$, $\frac{A_i}{\omega-z_i}$, with:
\begin{equation}
    \frac{A_i}{\omega-z_i} \to
    \frac{f_i A_i}{\omega-z_i}+
    \frac{(1-f_i) A_i}{\omega-z_i^*},
\end{equation}
where $f_i = f(\text{Re}z_i)$,  and $f(\omega)$ ($1-f(\omega)$) is the smearing function for the poles with $\text{Re}z_i \le \mu$ ($\text{Re}z_i > \mu$).

Importantly, this smearing preserves the zeroth total moment of the Green's function as defined in Ref.~\cite{chiarotti_unified_2022}---as the integration on a big circle encloses all the poles---i.e., the normalization of the spectral function, and the real part of the first total moment (this latter is true only if the amplitudes are real). 
Also, note that for a non-interacting system, this smearing procedure yields the same result of smearing the one-particle density matrix, as is typically done in KS-DFT implementations, i.e., $\gamma = \sum_i f_i \ket{\psi_i^\text{KS}}\bra{\psi_i^\text{KS}}$.

To reduce computational costs, it is sufficient to apply the smearing only within a sufficiently wide energy window near the chemical potential $\mu$, while leaving the rest of the SOP of $G$ unchanged.  
The width of this window is estimated at the beginning of the calculation by evaluating the particle number from $G^\text{KS}(\omega)$, considering the noninteracting chemical potential $\mu^\text{KS}$: if the particle number is accurate within numerical precision (typically $\sim 10^{-12}$), that window is used---bearing in mind that, during the self-consistent cycle, the window must remain large enough to accommodate potential metal-insulator transitions.

\begin{comment}
    \MQ{In passing, if you also take the adjoint $A^\dagger$ of the residue when doing the complex conjugation of the pole $z^*_i$, you get the following
\begin{align}
    b_i(\omega)&=\frac{A_i}{\omega-z_1}\\
    b_i^\text{smear}(\omega)&=\frac{f(z_i)A_i}{\omega-z_i}+\frac{(1-f(z_i))A_i^\dagger}{\omega-z_i^*}\\
    \text{Re}[b_i^\text{smear}(\omega)]&=\text{Re}[b_i(\omega)]\\
    \text{Im}[b_i^\text{smear}(\omega)]&=\text{Im}[b_i(\omega)]\times\left[2f(z_i)-1\right]\text{sgn}(\text{Im}[z_i]),
\end{align}
i.e. you alter the imaginary part of the SOP, which is related to the DOS, without altering the real part.
For matrix-valued residues real and imaginary part should be substituted by hermitian and nonhermitian part.
}
\end{comment}

\section{Self-consistent search for the chemical potential}
\label{sec:chem_pot}
At each iteration, the chemical potential $\mu$ is determined by imposing particle number conservation.
That is, $\mu$ is chosen such that $N = \int_{-\infty}^{\mu} \text{Tr}\, A(\omega)\, d\omega$.
Here we describe the procedure for smeared propagators.
Following the approach typically used in standard KS-DFT implementations, for a given $G(\omega)$ represented on a smeared SOP, we proceed self-consistently as follows:
\begin{enumerate}
    \item Set the chemical potential to two initial trial values $\mu_\text{trial}^{\pm}$ such that $\mu_\text{trial}^{-} < \mu_\text{trial}^{+}$;
    \item Using the procedure described in Appendix~\ref{sec:smearingSOP}, obtain the two corresponding smeared Green's functions on SOP, denoted $\mathcal{G}^{\pm}(\omega)$, for each $\mu_\text{trial}^{\pm}$;
    \item Compute the particle numbers $N^\pm_\text{trial}$ (the zeroth occupied moment of the Green's function) by summing over the occupied poles: $N^\pm_\text{trial} = \sum_i^{\text{Im}[z_s]>0} \text{Tr}[\mathcal{A}_i^\pm]$;
    \item Ensure that $N^{-}_\text{trial} < N < N^{+}_\text{trial}$, which implies 
    $\mu_\text{trial}^{-} < \mu < \mu_\text{trial}^{+}$;
    \item Apply a bisection algorithm, repeating the steps above and narrowing the interval to determine $\mu$ to machine precision.
\end{enumerate}

In addition to determining the chemical potential to preserve particle number, we introduce a small static imaginary correction to the self-energy in order to enforce the exact sign change of its imaginary part at the initial chemical potential $\mu_0$.
This is an exact constraint satisfied at self-consistency that we enforce here to improve the numerical stability of the calculations.

\renewcommand{\emph}{\textit}
\bibliographystyle{apsrev4-1}
\bibliography{references.bib,references_extra.bib}

\begin{thebibliography}{71}
\bibitem [{not()}]{notes_for_DOS}%
  \BibitemOpen
  \href@noop {} {}\bibinfo {note} {Note that, due to the finite lifetimes, we estimate the occupied bandwidth and effective mass, i.e. the position of the first $t_{2g}$ quasiparticle peak (BW) and last peak (full), as the center of the peak of the DOS. In the figure we do not use an extra broadening for the plotting.}\BibitemShut {Stop}%
\end{thebibliography}

\end{document}